\documentclass[twocolumn, twocolappendix]{aastex631}

\usepackage{multirow}
\usepackage{amsmath}
\usepackage{cases}

\begin{document}

\title{Dependence of Planet populations on Stellar Mass and Metallicity: A Pebble Accretion-based Planet Population Synthesis}

\correspondingauthor{Beibei Liu}
\email{bbliu@zju.edu.cn}

\author{Mengrui Pan}
\affiliation{Institute for Astronomy, School of Physics, Zhejiang University, Hangzhou 310027, China\\}
\affiliation{Center for Cosmology and Computational Astrophysics, Institute for Advanced Study in Physics, Zhejiang University, Hangzhou 310027, China\\}

\author{Beibei Liu}
\affiliation{Institute for Astronomy, School of Physics, Zhejiang University, Hangzhou 310027, China\\}
\affiliation{Center for Cosmology and Computational Astrophysics, Institute for Advanced Study in Physics, Zhejiang University, Hangzhou 310027, China\\}

\author{Linjie Jiang}
\affiliation{Institute for Astronomy, School of Physics, Zhejiang University, Hangzhou 310027, China\\}
\affiliation{Center for Cosmology and Computational Astrophysics, Institute for Advanced Study in Physics, Zhejiang University, Hangzhou 310027, China\\}

\author{Jiwei Xie}
\affiliation{School of Astronomy and Space Science, Nanjing University, Nanjing 210023,  China\\}
\affiliation{Key Laboratory of Modern Astronomy and Astrophysics, Ministry of Education, Nanjing 210023,  China\\}

\author{Wei Zhu}
\affiliation{Department of Astronomy, Tsinghua University, Beijing 100084, China\\}

\author{Ignasi Ribas}
\affiliation{Institut de Ci{\`e}ncies de l'Espai (ICE, CSIC), Campus UAB, c/ Can Magrans s/n, 08193 Bellaterra, Barcelona, Spain \\ }
\affiliation{Institut d'Estudis Espacials de Catalunya (IEEC), c/ Gran Capit{\`a} 2–4, 08034 Barcelona, Spain \\}


\begin{abstract}

The formation and evolution of planetary systems are linked to their host stellar environment. In this study, we employ a pebble accretion-based planet population synthesis model to explore the correlation between planetary properties and stellar mass/metallicity. Our numerical results reproduce several main aspects of exoplanetary observations. First, we find that the occurrence rate of super-Earths $\eta_{\rm SE}$ follows an inverted V-shape in relation to stellar mass: it increases with stellar mass among lower-mass dwarfs, peaks at early-M dwarfs, and declines toward higher-mass GK stars. Second, super-Earths grow ubiquitously around stars with various metallicities, exhibiting a flat or weak $\eta_{\rm SE}$ dependence on $Z_{\star}$. Third, giant planets, in contrast, form more frequently around stars with higher-mass/metallicity. Lastly, we extend a subset of simulations to $1$ Gyr to investigate the long-term evolution of the systems' architecture. By converting our simulated systems into synthetic observations, we find that the eccentricities and inclinations of single-transit systems increase with stellar metallicity, while these dependencies in multi-planet systems remains relatively weak. The alignment between our results and observations provides key insights into the connection between planet populations and stellar properties.

\end{abstract}


\section{Introduction} \label{sec:intro}

A significant number of exoplanets have been discovered around stars of various masses and metallicities, providing a robust dataset for statistical studies \citep{Zhu&Dong2021}. The planet occurrence rate $\eta_{\rm p}$, defined as the average number of planets per star ($N_{\rm p}/N_{\star}$), offers a quantitative measure of how frequently planets occur around stars within specific parameter spaces \citep{Mayor2011, Kopparapu2013, Petigura2013, Zhu2018, Hsu2019, Yang2020}.

Numerous studies have investigated the planet populations in relation to the mass $M_{\star}$ and  metallicity $Z_{\star}$ of their hosts \citep{Johnson2010, Buchhave2012, Schlaufman2015, Reffert2015, Ghezzi2018, Boley2024}. Observations reveal that super-Earths are more prevalent around lower mass stars; their occurrence rate anti-correlates with stellar mass \citep{Howard2012, Mulders2015b, Yang2020, He2021} and peaks at early-M dwarfs \citep{Sabotta2021}. However, giant planets are more frequently found around higher-mass stars, displaying a positive correlation between their occurrence rate and $M_{\star}$ \citep{Johnson2010, Gaidos2013, Fulton2021}. 

On the other hand, super-Earths and giant planets exhibit distinct dependencies of their occurrence rates on stellar metallicity \citep{Mayor2011}. Super-Earths are detected around stars spanning a broad range of metallicities, suggesting a weak or no correlation with $Z_{\star}$ \citep{Sousa2008, Buchhave2012, Schlaufman2015,Courcol2016, Petigura2018}, while giant planets are predominantly found around metal-rich stars \citep{Fischer&Valenti2005, Narang2018}.

The observed planet occurrence rates play a crucial role in constraining the theoretical planet formation models. The correlations of $\eta{-}M_{\star}$ and $\eta{-}Z_{\star}$ have been tested for both planetesimal and pebble accretion models \citep{Ida&Lin2004, Ida&Lin2005, Mordasini2009, Miguel2011, Ronco2017,  Liu2019, Burn2021, Emsenhuber2021, Boettner2024, Burn&Mordasini2024, Sanchez2024}. Based on the latest planetesimal population synthesis model, \citet{Burn2021} reproduced an increasing occurrence rate of giant planets with stellar mass and demonstrated the requirement of relatively high metallicity for giant planet formation. Their peak super-Earth occurrence rate near solar-type stars, however, appears to differ from observations.

From the pebble accretion perspective, \citet{Mulders2021} proposed that the decreased occurrence rate of transiting super-Earths around higher-mass stars can be explained by the growth of giant planets in their outer disk regions. 
Once the planet cores reach the pebble isolation mass \citep{Lambrechts2014}, the inward drift of pebbles gets terminated, suppressing the further growth of close-in super-Earths\footnote{Recent studies nevertheless proposed that the drifting pebbles cannot be entirely blocked by the massive planets, depending on disk turbulence \citep{Liu2022b, Stammler2023}. }. However, \citet{Chachan&Lee2023} suggested that sufficient material had already drifted into the inner disk prior to the formation of outer isolation-mass planets. Nevertheless, these models do not consider planet migration, which may play a crucial role in shaping the formation and radius distribution of super-Earths and sub-Neptunes (see \citealt{Venturini2024}). Furthermore, they only consider the growth of a single embryo. Including the disk migration and N-body effect is a natural extension for our study.



Here we employ a pebble accretion-driven planet population synthesis model to investigate multi-planet formation around stars with varying masses and metallicities. 
An overview of the synthesis model and the setup of initial conditions is introduced in Sect. \ref{sec:PPS}. The occurrence rates of different planet populations on stellar mass are demonstrated in Sect. \ref{sec:Occ}. We discuss the influence of stellar metallicity in Sect. \ref{sec:Metal}. Our findings are summarized in Sect. \ref{sec:sum}.

\section{Method} \label{sec:PPS}

In this section, we introduce the basics of our planet population synthesis model. The key planet formation processes and initial setups are described in Sect. \ref{sec:model} and Sect. \ref{sec:IC}, respectively.   

\subsection{Model description}
\label{sec:model}

Our framework is based on the latest pebble accretion-driven planet formation model of \citet{Liu2019} and \citet{Pan2024}.  The detailed descriptions are in Section 2 of \citet{Pan2024}. In this study, we have developed updated modules that incorporate the luminosity evolution of pre-main-sequence stars and chemical compositions for pebbles at different disk regions (see Appendices \ref{app:luminosity} and \ref{app:chemical}). 
We present an outline of the key physical processes included in our model as follows.

Following a conventional parametric 1D viscous $\alpha$-approach \citep{Shakura&Sunyaev1973}, we adopt a disk model composed of an inner viscously heated region and an outer stellar irradiation region \citep{Garaud&Lin2007, Ida2016,Liu2019}. Planets first grow by accreting pebbles \citep{Ormel&Klahr2010, Lambrechts2014, Liu&Ormel2018, Ormel&Liu2018} and then initiate runaway gas accretion once reaching the pebble isolation mass \citep{Johansen2019,Liu2019}. 
The planet with a mass higher than the isolation mass opens a gap in the disk, halting the inward drift of pebbles at the outer edge of the gap. Besides, stellar UV/X-ray photoevaporation can also carve a gap during the late stage of disk evolution, typically at a few Myr and at a distance of a few au, which may further limit the planet growth \citep{Venturini2020} and influence the subsequent dynamical evolution of system \citep{Liu2022}. The effect of stellar photoevaporation is not considered in this study.

On the other hand, low-mass planets undergo type I migration through the interaction with the disk gas, the direction and speed of which depends on the disk structure and thermal properties \citep{Paardekooper2011}.  
Recent studies have indicated that additional asymmetric torques arise when considering the energy release during solid accretion \citep{Benitez-Llambay2015, Masset2017, Guilera2021} or dust drifting flows \citep{Benitez-Llambay2018, Guilera2023, Chrenko2024}. These torques may alter the timescale and direction of migration. In this work, we adopt the type I torque based on the formulas of \cite{Paardekooper2011}, where outward migration occurs only in the inner viscously heated region. The potential implications of the other two torques will be investigated in future studies.
When the planets become massive enough to open a deep gap, they transition to the slower Type II migration  \citep{Kanagawa2015, Kanagawa2018}. Meanwhile, planets also experience collisions/scatterings as a result of gravitational interactions during close encounters \citep{Goldreich2004, Wimarsson2020}.

The Stokes number ($\rm St$) is the key dimensionless parameters that measure the aerodynamic size of pebbles. It also crucially influences the pebble accretion efficiency. In contrast to the assumption of a constant Stokes number in \citet{Pan2024}, we consider pebble growth to be limited by either fragmentation or radial drift.

In the fragmentation-limited regime, the maximum Stokes number is given by  \citep{Ormel&Cuzzi2007, Birnstiel2009}
\begin{equation}
    {\rm St}_{\rm frag} = \frac{1}{3 \alpha_{\rm t}} \frac{v_{\rm f}^2}{c_{\rm s}^2},
\end{equation}
where $\alpha_{\rm t}$ quantifies the disk midplane turbulent strength,  $c_{\rm s}$ is the gas sound speed, and $v_{\rm f}$ is the fragmentation threshold velocity that depends on the constitution of pebbles. Here we use $v_{\rm f} {= }1 \, \rm m \, s^{-1}$ interior to the water ice line and $v_{\rm f} {=} 10 \, \rm m \, s^{-1}$ exterior to it, reflecting the increased stickiness of icy grains compared to silicate grains \citep{Gundlach&Blum2015, Musiolik2021}.

In the drift-limited regime, the maximum Stokes number is determined by the balance between the radial drift and the growth \citep{Weidenschilling1977, Lambrechts&Johansen2014, Ida2016}, which can be expressed as
\begin{equation}
    {\rm St}_{\rm drift} = \frac{\sqrt{3 \pi}}{80} \, \frac{v_{\rm K}}{\Delta v} \, Z_0,
\end{equation}
where $v_{\rm K}$ is the Keplerian velocity, $Z_0$ is the initial metallicity of the disk, and $\Delta v {=} \eta v_{\rm K} {=} - \frac{1}{2} \, (\frac{H}{r})^2 \, \frac{\partial \ln P}{\partial \ln r} \, v_{\rm K}$ defines the deviation between the gas azimuthal and Keplerian velocities. The actual Stokes number is then calculated as 
\begin{equation}
    {\rm St} = \rm min({\rm St}_{\rm frag}, \ {\rm St}_{\rm drift}).
\end{equation}
This approach allows us to model the pebble size more realistically. 

We assume the dust-to-gas ratio is \citep{Burn2021}
\begin{equation}
    \xi = \frac{\dot{M}_{\rm peb}}{\dot{M}_{\rm g}}= 10^{\rm Fe/H} \, f_{\rm dg, \odot} \, Cf_{\rm mass},
\end{equation}
where $f_{\rm dg, \odot} = 0.0149$, $Cf_{\rm mass}$ is the cumulative mass fraction of different chemical species (see Table \ref{tab:species}).
It is worth noting that more sophisticated dust growth and evolution models yield a more complex, time and radius dependent pebble flux \citep{Mulders2021, Drazkowska2021, Fang2023}. Moreover, given that the pebble drift velocity varies radially with the Stokes number, the solution of the advection-diffusion equation predicts a traffic-jump effect near the ice line \citep{Venturini2020, Guilera2021}. This effect, neglected in our study,  may result in an enhancement of the pebble flux in the inner disk, further facilitating the formation of planets in this region. 

\subsection{Initial condition setup}
\label{sec:IC}
\begin{table*}[]
    \centering
    \caption{ Monte Carlo sampling of initial parameters and the number of simulations in our study.}
    \begin{tabular}{ccccccc}
        \hline
        \hline
        $M_{\star}$ & $L_{\star, 0}$ & $\dot{M}_{\rm g, 0}$ & [Fe/H] &  $P_{\star, \rm rotate}$ & $\alpha_{\rm t}$ & $N_{\rm sim}$\\
        ($M_{\odot}$) & ($L_{\odot}$) & ($M_{\odot} \ \rm yr^{-1}$ ) & & (days) &  &\\
        \hline
        0.1 & $\log\mathcal{N}(-1.16, \, 0.17^2)$ & $\log\mathcal{N}(-9.44, 0.74^2)$ & \multirow{4}{*}{$\mathcal{N}(-0.07, \, 0.21^2)$} &  \multirow{4}{*}{$\log\mathcal{N}(4.74, \, 2.02^2)$}  & \multirow{4}{*}{$\log\mathcal{U}[10^{-4}, \, 5 \times 10^{-3}]$} & 559\\
        0.3 & $\log \mathcal{N}(-0.52, \ 0.17^2)$ & $\log\mathcal{N}(-8.59, 0.93^2)$ & &   &  & 255 \\
        0.5 & $\log\mathcal{N}(-0.25, \ 0.17^2)$ & $\log\mathcal{N}(-8.19, 0.89^2)$ &  &  &  & 155\\
        1.0 & $\log\mathcal{N}(0.29, \ 0.17^2)$ & $\log\mathcal{N}(-7.83, 0.82^2)$ &  &   &  & 225\\
        \hline
    \end{tabular}
    \label{tab:setup}
\end{table*}

We employ the MERCURY N-body code \citep{Chambers1999} to investigate the formation and evolution of planet systems by sampling the initial conditions using a Monte Carlo approach. The Monte Carlo variables include the initial stellar luminosity $L_{\star, 0}$, the initial disk accretion rate $\dot{M}_{\rm g, 0}$, the stellar metallicity $[\rm Fe/H]$, the inner disk edge $r_{\rm in}$, and the disk turbulent parameter $\alpha_t$. We vary the stellar mass from $0.1 $, $0.3$, $0.5$ to $1 \ M_{\odot}$.  Table~\ref{tab:setup} summarizes the corresponding parameter distributions and the number of simulations conducted in this study.

We adopt a log-normal distribution for initial stellar luminosity, with a mean value from  \citet{Baraffe2015}'s evolutionary models and a spread of $\sigma {= }0.17$ dex to account for stochastic variations in the stellar formation environment. The initial disk accretion rate follows a log-normal distribution based on the measurements from different star-forming region, including Lupus \citep{Alcala2014, Alcala2017}, Chamaeleon I \citep{Manara2016, Manara2017}, Chamaeleon II \citep{Villenave2021} , and $\beta$-Ophiuchus \citep{Testi2022}. We refer to Table~1 of \citet{Manara2023} for the corresponding data. 
As the same in \cite{Pan2024}, we use a simple accretion rate evolution such that  
\begin{subnumcases} {\dot{M}_{\rm g} = }
\dot{M}_{\rm g, 0} & for $t \leq t_{0}$ \\
\dot{M}_{\rm g, 0} \ \rm{exp}[-(t-t_0)/\tau_{\rm dep}] & for $t > t_0$. 
\end{subnumcases}Considering that the typical disk lifetime is ${\sim}2{-} 5$ Myr, for simplicity we assume that disk dissipation begins at $t_0 {= }1$ Myr, with a dispersal timescale $\tau_{\rm dep}$ of $0.5$ Myr.

The distributions of stellar metallicity are assumed to be the same among different stellar masses. Inferred from spectroscopic surveys of solar neighbor stars, we model $Z_{\star}$ following a Gaussian distribution with a mean of $-0.07$ and a standard deviation of 0.21 \citep{Santos2003, Burn2021}. The inner edge of the disk is set by the stellar corotation radius,  derived from the rotation periods of young T Tauri stars \citep{Venuti2017}. Here we adopt a log-normal rotation period distribution with a mean of 4.74 days and a spread of 2.02 days. 

Protoplanetary disks are expected to be turbulent, with the strength varying across systems \citep{Rosotti2023}. To encompass this broad parameter space and account for potential variability, we adopt a log-uniform distribution for the turbulent viscosity parameter $\alpha_{\rm t}$ spanning  $10^{-4}$ to $5 \times 10^{-3}$.  

We initially place $20$ protoplanetary embryos with a mass of $0.01 \ M_{\oplus}$ between $0.2 \, r_{\rm trans}$ and $5 \, r_{\rm trans}$, where $r_{\rm trans}$ is the disk radius that separates the inner viscously heated region from the outer stellar irradiated region\footnote{For instance, for a solar-mass star and disk accretion rate of $ 1.5 \times 10^{-8} \, M_{\odot} \, \rm yr^{-1}$,  $r_{\rm tran} \simeq 3.3 \, \rm au$.}. 
We do not model the detailed dust coagulation and planetesimal formation. All the planetary embryos are implemented at the start of the simulations (but see \citealt{Voelkel2022} for a different approach). The work focuses on the subsequent embryo growth, migration and dynamics.

Each simulation was integrated up to 5 Myr, by which time the disk had fully dissipated. 
The occurrence rates of various planets at this stage were compared with observations. On average, such a simulation required approximately three weeks of computational time on a single CPU.  In Sect. \ref{sec:ecc_inc}, we extended a subset of simulations to $10^9$ yr to investigate the long-term orbital evolution of planetary systems. These extended simulation requires three months to complete. This allowed us to compare dynamical indicators of the systems—such as planetary eccentricities and inclinations—with observations.

We note that several additional long-term effects, such as stellar tide,  UV/X-ray photoevaporation and core-power induced mass-loss are not included in our model. These processes could significantly influence the radii and dynamics of planets with shortest orbital periods \citep{Mardling&Lin2002, Owen&Wu2013, Wang&Lin2023}. Therefore, planets with orbital period $P$ ${<} 10$ days are excluded from our follow-up comparison, which means that we only focus on warm and cold populations of planets. Besides, the planet's mass is the quantity obtained from simulations. 
Our results can be directly compared with radial velocity observations. For transiting surveys, planet masses are inferred from their observed radii using the mass-radius relation from \cite{Chen&Kipping2017}.

\section{Dependency of planet occurrence rate on stellar mass} \label{sec:Occ}
 
In order to compare with observations more specifically, we divide the simulated planets into three groups: Earth-like and super-Earth planets ($1{-}10 \ M_{\oplus}$), Neptunian-mass and sub-giant planets ($10{-}100 \ M_{\oplus}$), and gas giant planets (${>} 100 \ M_{\oplus}$).  

Since the warm populations of planets currently have the most complete demographic statistics, we compare the occurrence rate of the three aforementioned planet populations in the orbital period range between $10$ and $100$ days in Sections \ref{sec:warmSE}, \ref{sec:warmN} and \ref{sec:warmG}, respectively.  We discuss the gas giant planets with $a{>1}$ au in Sect. \ref{sec:cold}. 
A detailed comparison of the planet mass and semimajor axis between our synthetic population and observations is presented in Appendix \ref{app:M_a_diagram}.


\begin{figure}
    \centering
    \includegraphics[width=\columnwidth]{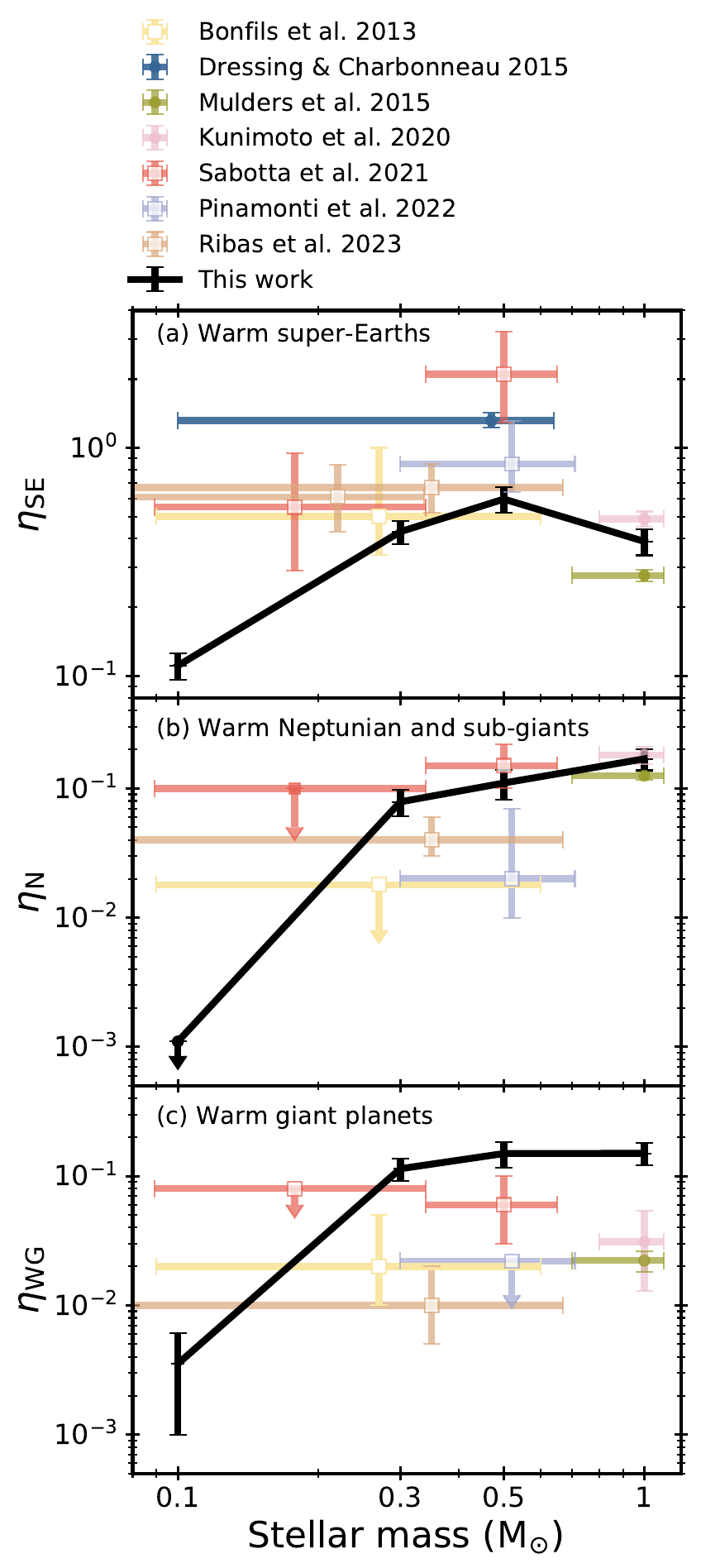}
    \caption{Comparison of the planet occurrence rate between population synthesis (in black) and observations (RV data in empty squares and Transit data in solid dots). Panel (a)-(c) show the occurrence rates of Super-Earths ($1-10 \ M_{\oplus}$ for RV or $1-2.8 \ R_{\oplus}$ for Transit data based on the Mass-Radius relation from \citet{Chen&Kipping2017}), Sub-Giants ($10-100 \ M_{\oplus}$ or $2.8-8 \ R_{\oplus}$), and Giant planets ($100-1000 \ M_{\oplus}$ or $8-16 \ R_{\oplus}$) within $10-100$ days as a function of stellar mass, respectively. Except for the occurrence rate of Super-Earths, which peaks near $0.5 \ M_{\odot}$ stars, the occurrence rates of Sub-Giants and Giants increase with stellar mass.}
    \label{fig:Occ_vs_star}
\end{figure}

\subsection{Warm Earths and super-Earths}
\label{sec:warmSE}
The occurrence rate of Earth and super-Earth planets can be given by $\eta_{\rm SE} {=} N_{\rm SE}/N_{\star}$, where $N_{\rm SE}$, and $N_{\star}$ are the numbers of super-Earths within the specified period and mass range and total stars, respectively.
Fig.~\ref{fig:Occ_vs_star}a shows $\eta_{\rm SE}$ for both simulations (black) and observations (colored), where transit and RV surveys are depicted as filled circles and open squares, respectively. 
For simulations, the vertical errorbar indicates the standard error from Poisson counting, while the horizontal colored bar refers to the range of stellar masses in the observed sample.

Previous studies have demonstrated considerable variation in $\eta_{\rm SE}$. For low-mass stars of $M_{\star}{\leqslant} 0.3 \ M_{\odot}$, \citet{Bonfils2013} and \citet{Sabotta2021} derived similar $\eta_{\rm SE}$ of $0.52_{-0.16}^{+0.50}$ and $0.55_{-0.26}^{+0.40}$ based on different RV samples. However, differed from RV results, $\eta_{\rm SE}$ for $1{-}2.5 \ R_{\oplus}$ planets obtained from the Kepler survey is systematically higher, likely due to a higher median mass of their stellar sample \citep{Dressing&Charbonneau2015}. Our simulations reveal that $\eta_{\rm SE}$ for stars of $M_{\star}{=}0.1 \ M_{\odot}$ is $0.11 \pm 0.01$. This value increases to $0.43 \pm 0.05$ for stars of $M_{\star}{=}0.3 \ M_{\odot}$, in line with the RV results.

For early M dwarf stars, various surveys suggested  $\eta_{\rm SE}$ ranging from 0.5 to 3 \citep{Dressing&Charbonneau2015, Mulders2015, Pinamonti2022, Ribas2023}. Our simulation gives $\eta_{\rm SE}{=}0.60 \pm 0.08$. 
For solar-mass stars, the observed $\eta_{\rm SE}$ declines to  ${\sim}0.26{-}0.5$ \citep{Mulders2015, Kunimoto&Matthews2020}. This, again, is comparable to our simulation of $0.39 \pm 0.05$. 

The inverted V-shape of $\eta_{\rm SE}$ as a function of stellar mass shown in  Fig.~\ref{fig:Occ_vs_star}a can be explained by several reasons. First, disks around more massive stars contain a higher amount of solids, facilitating the formation of protoplanetary cores. On the other hand, disks around more massive stars could be generally larger \citep{Bate2018}. As stellar mass increases further, although super-Earths may form more readily, the larger disk sizes favor planet formation in the extended outer regions \citep{Pan2024}. Given that the type I migration timescale scales with the star-to-planet mass ratio \citep{Tanaka&Ward2004}, super-Earths migrate more slowly around high-mass stars compared to those around low-mass M dwarfs \citep{Alibert&Benz2017, Miguel2020, Burn2021, Gan2024, Johnston2024, Venturini2024}. Since we focus on the warm super-Earth population, embryos form at wider orbital distances and later undergo slower disk migration, leading to a lower $\eta_{\rm SE}$ at short orbital distances around more massive stars.  

Furthermore, pebble accretion efficiency scales with planet Hill radius, which is larger around lower-mass stars \citep{Liu&Ormel2018, Ormel&Liu2018}. This also explains the decrease of $\eta_{\rm SE}$ from early M dwarfs to solar-mass stars. Besides, solar-type stars are more likely to form massive planets. On the one hand, these planets can act as barriers to impede the inward transportation of solids (either pebbles or planets) to the inner disk regions \citep{Ormel2017, Izidoro2015, Lambrechts2019}. On the other hand, even the formation of super-Earths remains unaffected by giant planets, these giant planets are strong perturbers, which may trigger dynamical instabilities within the system and hindering the survival of the inner super-Earths.  These two factors further contribute to the lower $\eta_{\rm SE}$ around solar-mass stars.

To summarize, $\eta_{\rm SE}$ increases with stellar mass from late to early M dwarfs, peaks at $M_{\star}{\simeq}0.5 \ M_{\odot}$ stars and declines as stellar mass further increases. The initial rise is primarily due to the higher solid mass budget around early M dwarfs, while the subsequent decrease results from a combination of factors, including disk size, disk migration, and the presence of giant planets.

\subsection{Warm Neptunian planets and sub-giants}
\label{sec:warmN}

Different from the trend for $\eta_{\rm SE}$, the occurrence rate of warm Neptunian planets and sub-giants $\eta_{\rm NSG}$ increases monotonically with stellar mass (Fig.~\ref{fig:Occ_vs_star}b). These intermediate-mass planets are seldom detected around stars of $M_{\star}{\leqslant} 0.3 \ M_{\odot}$ \citep{Bonfils2013, Sabotta2021}. From simulations we find an upper limit of $\eta_{\rm NSG}{=}0.1\%$ for stars of $M_{\star}{=}0.1 \ M_{\odot}$ and ${=}0.08 \pm 0.02$ for stars of $M_{\star}{=}0.3 \ M_{\odot}$. These are consistent with the upper limit estimates from  \cite{Sabotta2021}.

After a nearly two-order-of-magnitude increase of $\eta_{\rm NSG}$ for stars from $0.1$ to $0.3 \ M_{\odot}$, the rise becomes slower towards higher mass stars. Averagely, $\eta_{\rm NSG}{=}0.11 \pm 0.03$ for stars of  $M_{\star}{=}0.5 \ M_{\odot}$ and $0.17 \pm 0.03$ for solar-mass stars. Our result broadly agrees with both RV and transit surveys \citep{Mulders2015, Kunimoto&Matthews2020, Sabotta2021, Pinamonti2022}.

\subsection{Warm giant planets} \label{sec:warmG}

More than a hundred warm giant planets have been detected, most of which orbit solar-like stars. Generally, the occurrence rate of warm gas giant planets $\eta_{\rm WG}$, is approximately $0.02{-}0.03$ for solar-mass stars \citep{Mulders2015, Kunimoto&Matthews2020}. Due to the lack of detections in the very low-mass stellar sample of the CARMENES project \citep{Sabotta2021}, only an upper limit has been derived. Mid-to-early M dwarfs host an average of $0.01{-}0.02$ warm Jupiters \citep{Pinamonti2022, Ribas2023}. For early M dwarfs, \citet{Sabotta2021} estimated a relatively higher $\eta_{\rm WG}$ compared to \citet{Ribas2023}. This difference is likely due to more efficient planet detection in the pre-selected CARMENES targets.

In our simulations, only one out of more than five hundred systems around stars of $M_{\star}{=}0.1 \ M_{\odot}$ contains a giant planet, indicating an extremely low $\eta_{\rm WG}$ of $0.35 \pm 0.25\%$. However, for more massive stars with $M_{\star}{=}0.3{-}1 \ M_{\odot}$, we find a slightly higher occurrence rate compared to observations.

This discrepancy between simulations and observations may be caused by two reasons. First, relatively fast   inward migration increases the fraction of warm gas giants. We follow the type II migration prescription of \citet{Kanagawa2018}, in which giant planets only migrate inward.  However, several studies have challenged this simplified picture. For instance, \citet{Liu2022} proposed that one-sided torque at the inner disk edge can drive outward migration during gas disk dispersal. Based on hydrodynamic simulations, \citet{LiYP2024} found that the migration direction of accreting giant planets can be reversed due to asymmetric spiral arms feeding from the global disk into its Hill radius. Hence, migration can be more complicated than the treatment we have adopted here.

Second, multiple giant planets can form early during the disk's lifetime, leading to a relatively high initial multiplicity. After the disk disperses, these giant planets may trigger dynamical instabilities. Planet-planet scatterings, collisions, and ejections would then reduce the multiplicity \citep{Ford&Rasio2008, Marzari2010}. This process could explain the lower $\eta_{\rm WG}$ observed in mature planetary systems compared to our synthetic populations at the end of the formation epoch.

\subsection{Cold giant planets}
\label{sec:cold}

\begin{figure}
    \centering
    \includegraphics[width=\columnwidth]{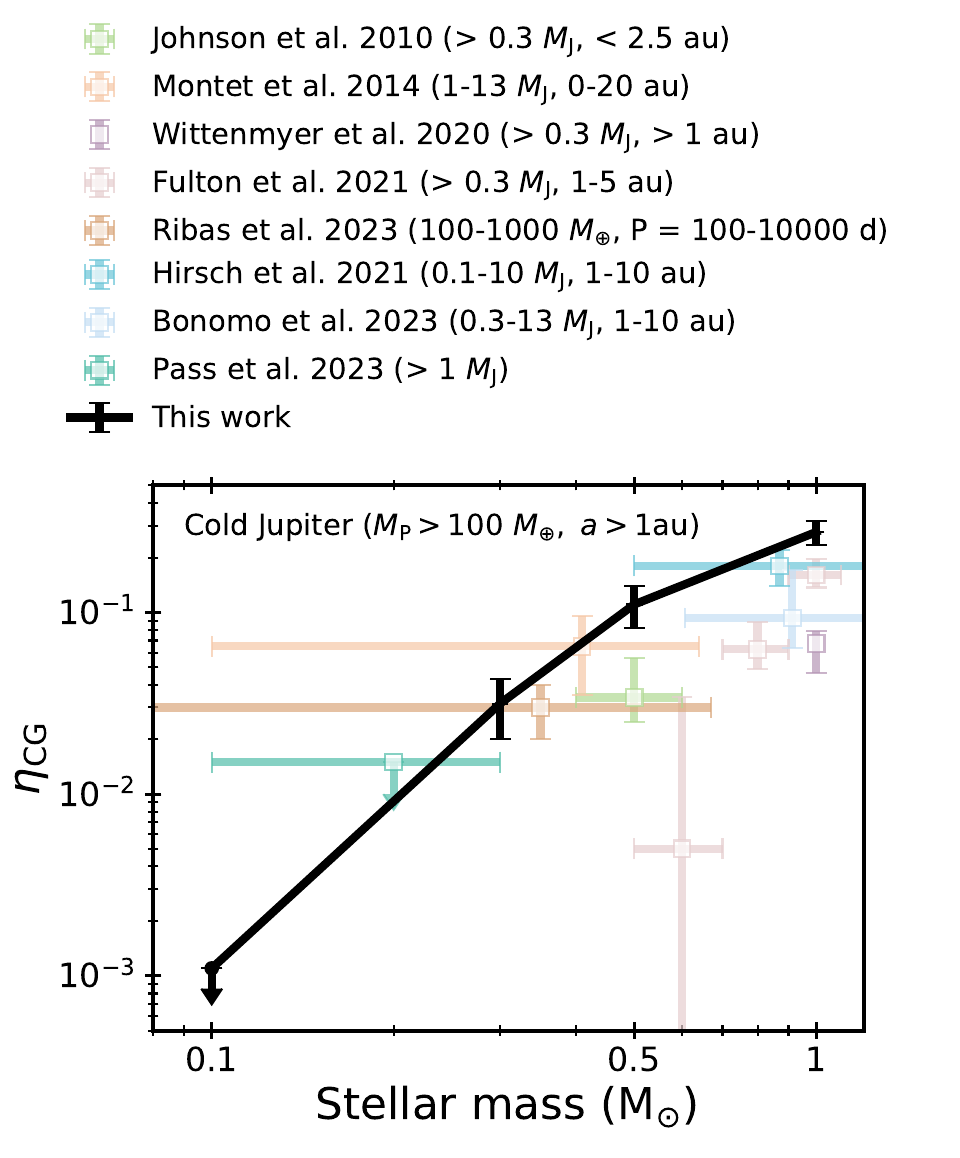}
    \caption{Occurrence rate of cold giant planets ($\eta_{\rm CJ}$), defined as planets with masses greater than $100 \ M_{\odot}$ and semi-major axis $a > 1 \ au$, increases monotonically with stellar mass. Specifically, $\eta_{\rm CJ} \leqslant 0.1\%$ around $0.1 \ M_{\odot}$ stars, while it is $0.03 \pm 0.01$, $0.11 \pm 0.03$, $0.28 \pm 0.04$ for stars with masses of $0.3 \ M_{\odot}$, $0.5 \ M_{\odot}$, and $1.0 \ M_{\odot}$, respectively.}
    \label{fig:Occ_CJ}
\end{figure}

The cold giant planets with orbital distances ${\gtrsim}$ 1 au is another population whose occurrence rate $\eta_{\rm CG}$ has been relatively well constrained. \citet{Pass2023} reported a null detection of giant planets around M dwarfs of $M_{\star}{=}0.1{-}0.3 \ M_{\odot}$ within 15 pc, providing a 95\% confidence upper limits of $ \eta_{\rm CG}{=}1.7\%$  ($M_{\rm p}{\sim}0.8{-}3 \ M_{\rm J}$ and orbital distance beyond the water ice line). Based on the California Planet Survey, \citet{Johnson2010} obtained the occurrence rate of giant planets ($M_{\rm p} \sin i {>} 100 \ M_{\oplus}$) of $3.4_{-0.9}^{+2.2}\%$ within 2.5 au around M dwarfs of $M_{\star}{< }0.6 \ M_{\odot}$. Combining RV and direct imaging measurements, \citet{Montet2014} inferred $\eta_{\rm CG}{=}6.5\% \pm 3.0\%$ around M dwarfs ($1{-}13 \ M_{\rm J}$ within 20 au). Besides, an occurrence rate of $3.0_{-1.0}^{+1.0}\%$ for giant planets with periods of $100{-}1000$ days was derived from the CARMENES survey \citep{Ribas2023}. 

At higher stellar masses, \citet{Hirsch2021} reported an occurrence rate of $0.18_{-0.03}^{+0.04}$ for giant planets with masses of $0.1{-}10 \ M_{\rm J}$ and orbit within $0.1{-}10$ au. \citet{Fulton2021} found that $\eta_{\rm CG}$s ( $M_{\rm p}{>} 100 \ M_{\oplus}$ within $1{-}5$ au) are ${\sim}0.50_{-0.50}^{+2.93}\%$, $6.26_{-1.40}^{+2.57}\%$, $16.04_{-2.34}^{+3.69}\%$ for stars of $M_{\star}{=}0.6 \ M_{\odot}$, $0.8 \ M_{\odot}$ and solar-mass, respectively. Consistently \citet{Bonomo2023} and \citet{Wittenmyer2020}  derived $\eta_{\rm CG}$ to be $9.3_{-2.9}^{+7.7}\%$ and $6.73_{-1.13}^{+2.09}\%$ for cold Jupiters around solar-type stars, based on observations from the HARPS-N and Anglo-Australian Planet Search programs.

Our simulations exhibit a significant rise of $\eta_{\rm CG}$ with stellar mass. Specifically, we find an upper limit of 0.1\% for stars with $0.1 \ M_{\odot}$, while  $0.03 \pm 0.01$, $0.11 \pm 0.03$ and $0.28 \pm 0.04$ for stars with masses of $0.3 \ M_{\odot}$, $0.5 \ M_{\odot}$ and solar-mass, respectively. 

Overall, our results show good agreement with the observed $\eta_{\rm CG}$ but tend to overestimate $\eta_{\rm WG}$. The number of giant planets per system at the end of gas-disk dispersal phase ($t{=}5$ Myr) in our simulation is $2.82 \pm 0.40$, nearly twice the intrinsic multiplicity inferred from the California Legacy Survey \citep{Zhu2022}. We expect that this value will decrease through planet-planet scatterings during long-term evolution in gas-disk free phase. The preliminary results from extended simulations in Sect \ref{sec:ecc_inc} imply an insufficient reduction to fully match the observations. Additional mechanisms, such as external flyby stars in the cluster environment, may be needed to further destabilize the systems (see, e.g., \citealt{Cai2017}).

\section{DEPENDENCY OF PLANET OCCURRENCE
RATE ON STELLAR Metallicity} \label{sec:Metal}
In this section,  we discuss the influence of $Z_{\star}$ on super-Earth and giant planet occurrence rates in Sect. \ref{sec:occ_meta}, and the architecture of planet systems in Sect. \ref{sec:ecc_inc}.

\subsection{Occurrence rate} 
\label{sec:occ_meta}

\begin{figure}
    \centering
    \includegraphics[width=\columnwidth]{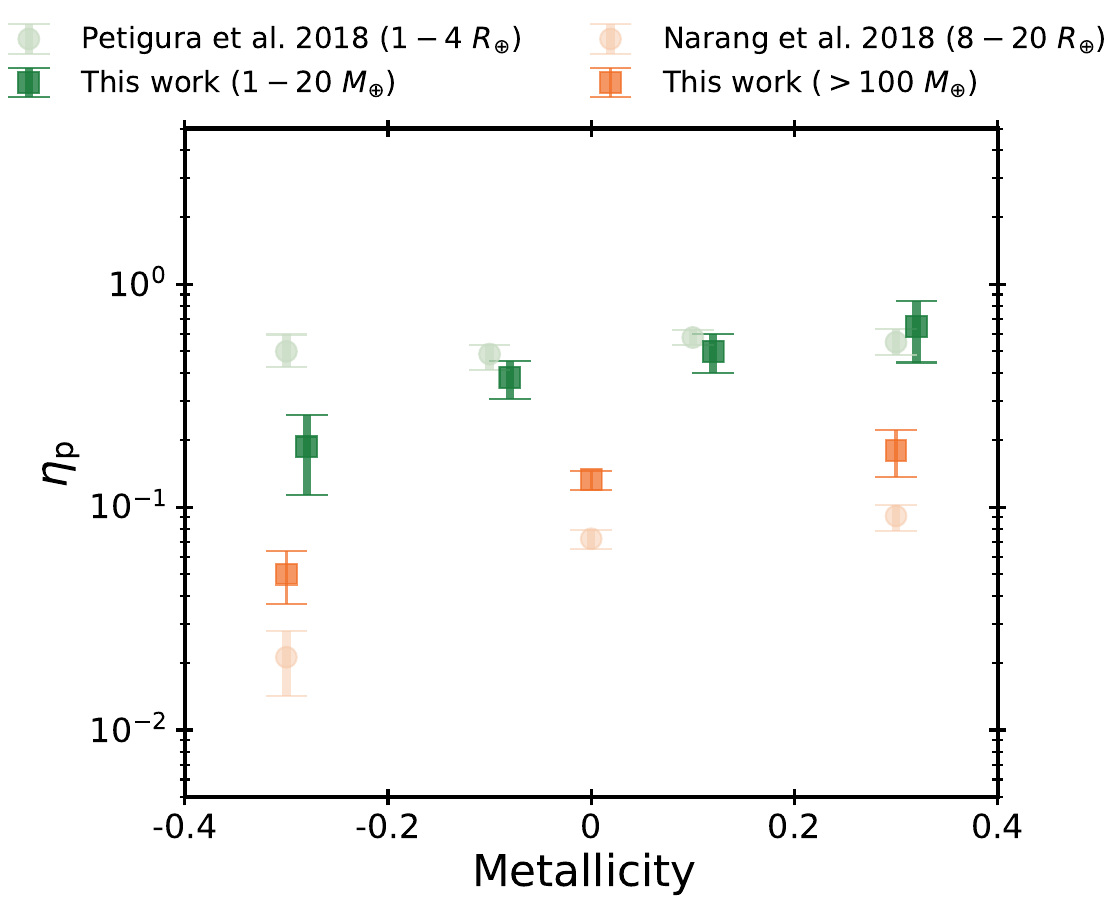}
    \caption{Occurrence rates between synthetic simulations and observations as a function of stellar metallicity. The thick green symbols represent the occurrence rates for simulated super-Earths ($M_{\rm p}{=}1{-}20 \ M_{\oplus}$ and $P {=} 10{-}100 \ \rm{days}$), whereas the light green indicates that of the observed planets with the same period range and $R_{\rm p}{=}1{-}4 \ R_{\oplus}$ from \citet{Petigura2018}. The thick and light orange symbols show the occurrence rates of simulated giant planets ($M_{\rm p}{>} 100 \ M_{\oplus}$, $P {=} 10-365 \ \rm{days}$) and \cite{Narang2018}'s sample from Kepler DR25 ($R_{\rm p}{=}8-20 \ R_{\oplus}$, $P {=} 1-365 \ \rm{days}$), respectively. The super-Earth occurrence rate remains nearly constant (or increases weakly) across the metallicity range, while that of the giant planets increases rapidly with $Z_{\star}$.}
    \label{fig:Occ_vs_Z}
\end{figure}

The positive correlation between giant planet occurrence rate and host star metallicity (hereafter $\eta_G{-}Z_\star$) strongly supports the core accretion theory (e.g., \citealt{Fischer&Valenti2005, Ida&Lin2004}).

Building upon the methodology of \citet{Narang2018}, we calculate the occurrence rate of giant planets (defined as $M_{\rm p}{>} 100 \ M_{\oplus}$ and orbital periods of $10{-}365$ days) as a function of $Z_{\star}$. The results are illustrated in Fig.~\ref{fig:Occ_vs_Z}, where our simulations and \citet{Narang2018}'s observation are marked as thick and light oranges, respectively. Although our simulation gives an occurrence rate slightly higher than that of \citet{Narang2018} (as discussed in Sect. \ref{sec:cold}), both studies reveal a similar trend of increasing occurrence rate with stellar metallicity.

Unlike giant planets,  low-mass planets show a flatter $\eta_{\rm SE}{-}Z_{\star}$ correlation. \citet{Buchhave2012} examined a sample of Kepler planet-hosting stars and found that small planets form around stars with a wide range of metallicities, which implies an overall weak or non-dependence on stellar metallicity (also see \citealt{Schlaufman2015}). Compared to metal-poor stellar sample, \citet{Wang&Fischer2015} found enhancement factors of $1.72^{+0.19}_{-0.17}$ and $2.03^{+0.29}_{-0.26}$ in metal-rich sample for the occurrence rates of terrestrial planets ($\leq 1.7 \ R_{\oplus}$) and sub-Neptunes ($1.7{-}3.9 \ R_{\oplus}$), respectively.  
Recently, \citet{Petigura2018} used the California-Kepler Survey (CKS) and reported a positive correlation for sub-Neptunes ($R_{\rm p}{=}1.7{-}4 \ R_{\oplus}$ and $P{<}100 \ \rm $ days), but no correlation for super-Earths ($R_{\rm p}{=}1{-}1.7 \ R_{\oplus}$) with a similar orbital period range. The situation could be different if the fraction of stars with planets is used to quantify the metallicity dependence \citep{Zhu2019}.

In Fig.~\ref{fig:Occ_vs_Z} we compare the simulations of small planets ($1{-}4 R_{\oplus}$) (thick green) with those of \citet{Petigura2018} (light green). A relatively flat occurrence rate is observed across metallicity in the range of $-0.2$ to $0.4$. The occurrence rate at very low metallicity ($Z_{\star} {<}-0.2$) shows a slight decrease compared to the observation. This discrepancy arises from the limited supply of pebbles in very metal-poor disks, where the mean solid mass is approximately $37 \ M_{\oplus}$ around solar-mass stars with $Z_{\star} {=} -0.3$.  Super-Earths are thus more challenging to form around very metal-poor stars. Our model only considers a smooth gas disk profile, in line with the recent observations showing that two-thirds of the samples in the Lupus star forming region are such disks \citep{Guerra-Alvarado2025}. However, planet growth in a structured disk remains a hot topic \citep{Guo2025}, as pebble accretion can be significantly enhanced by efficient pebble accumulation at pressure traps or the magnetospheric truncation radius \citep{Guilera2020, Morbidelli2020, Chambers2021, Li2022, Jiang&Ormel2023, Li2024}. This represents a promising avenue for future exploration.

\subsection{Eccentricity and inclination} \label{sec:ecc_inc}

\begin{figure*}
    \centering
    \includegraphics[width=\textwidth]{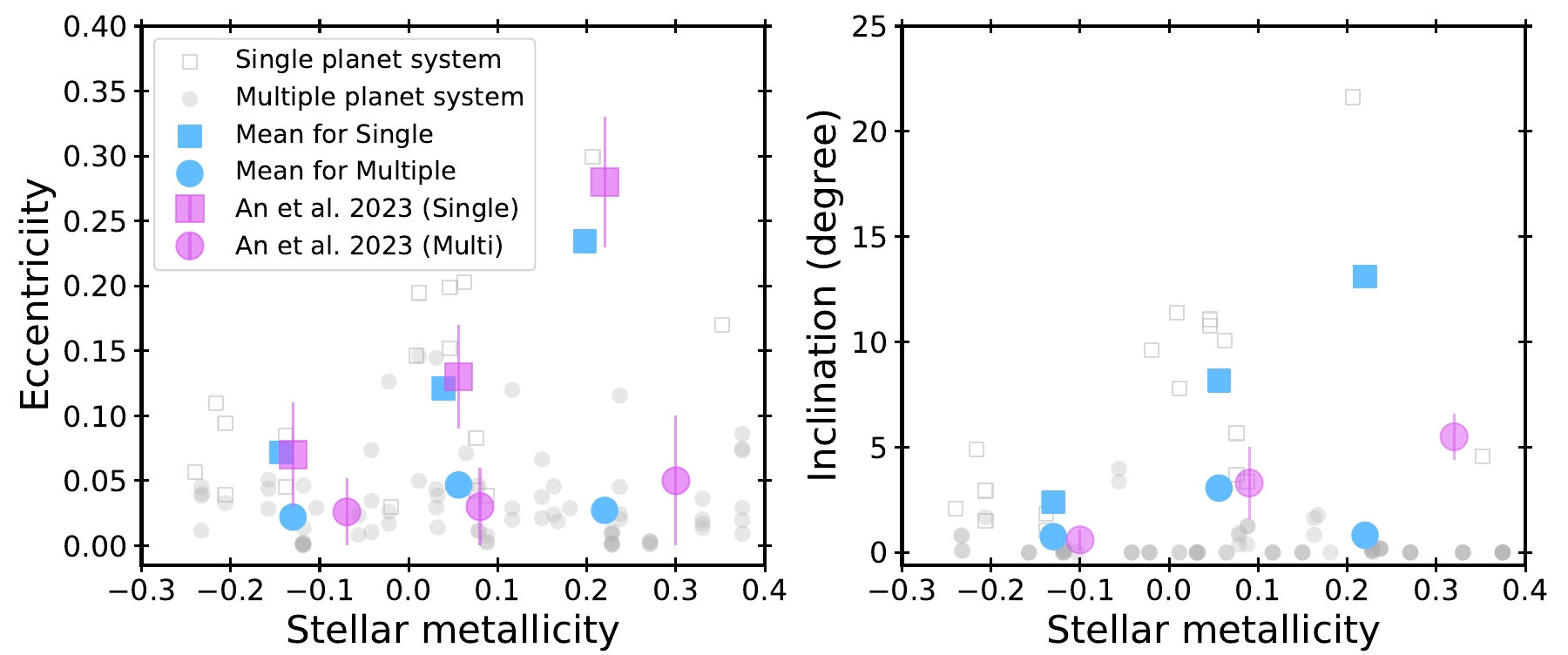}
    \caption{
    Orbital eccentricity (left) and inclination (right) between synthetic long-term simulations and observations as a function of stellar metallicity. The squares and circles represent the transiting singles and multiples.  The grey, blue and purple refer to individual simulation, the mean value among simulations in each metallicity bin and \cite{An2023}'s observation, respectively. Multi-planets formed around more metal-rich stars are likely to undergo more frequent dynamical instabilities, and therefore, the orbital eccentricities and inclinations of surviving planets increase with stellar metallicity. The systems that seldom undergo strong planet-planet scatterings remain stable and those planets are on nearly circular and coplanar orbits.
     }
    \label{fig:ecc_inc}
\end{figure*}

The correlation between the metallicity of host stars and the architecture of planet systems is a subject under active investigation. \citet{Santos2003} first suggested that eccentric planets tend to orbit predominantly around metal-rich stars. 
Then, \citet{Xie2016} discovered that the single transiting planets observed by Kepler exhibit higher eccentricities compared to multiple transiting planets, a finding subsequently validated by \citet{VanEylen2019} and \citet{Mills2019}.  While \citet{VanEylen2019} found no clear correlation between stellar metallicity and orbital eccentricity, \citet{Mills2019} tentatively suggested a preference for high metallicity in small eccentric planets. 
Lately, \citet{An2023} and \citet{Hua2025} confirmed the existence of eccentricity (inclination) and metallicity correlation, respectively. Their finding implies that stellar metallicity plays a vital role in shaping the planet system architectures \citep{Mishra2023}.

To explore this correlation in details, we extended the simulation integration time to $10^9$ yrs for dozens of systems that consist of only small planets around stars of various $Z_{\star}$. 
To account for the observational selection effect, we define the simulated single-transiting planet when fulfilling the criterion of $2R_{\star}/a_{\rm p} {<} \Delta i$, where $R_{\star}$ is the stellar radius, $a_{\rm p}$ is the planet semi-major axis and $\Delta i$ is the minimum mutual inclination between this planet and others in the system. 
To compare the synthetic mature population with those from \citet{An2023}, we follow their methodology and divide our simulations into three metallicity bins: $Z_{\star}<-0.05$, $-0.05 \leqslant Z_{\star} \leqslant 0.14$, and $Z_{\star}>0.14$. The results are illustrated in Fig.~\ref{fig:ecc_inc}, where the square and circle represent transiting single and multiple systems, and the grey, blue and purple symbols refer to individual simulation, mean value among simulations in each bin and observation, respectively.

As can be seen from Fig.~\ref{fig:ecc_inc}, the mean eccentricity and inclination of single-transiting-planet systems from our simulations (blue squares) exhibit an increasing trend with stellar metallicity, while multi-planet systems tend to remain nearly circular orbits (blue circles), consistent with \citet{An2023}'s result (purple). However, we obtain slightly lower mean values compared to \citet{An2023} at $Z_{\star}{\approx}0.2{-}0.3$. This might be partly due to the scarcity of systems around stars of high metallicity in our simulations. Multiple giant planets more frequently form in such high metallicity environments. These systems become long-term unstable, which impedes the late growth and survival of small planets. As a result, the total number of simulations we can have at such super-solar metallicity is largely reduced.

\begin{figure}
    \centering
    \includegraphics[width=\columnwidth]{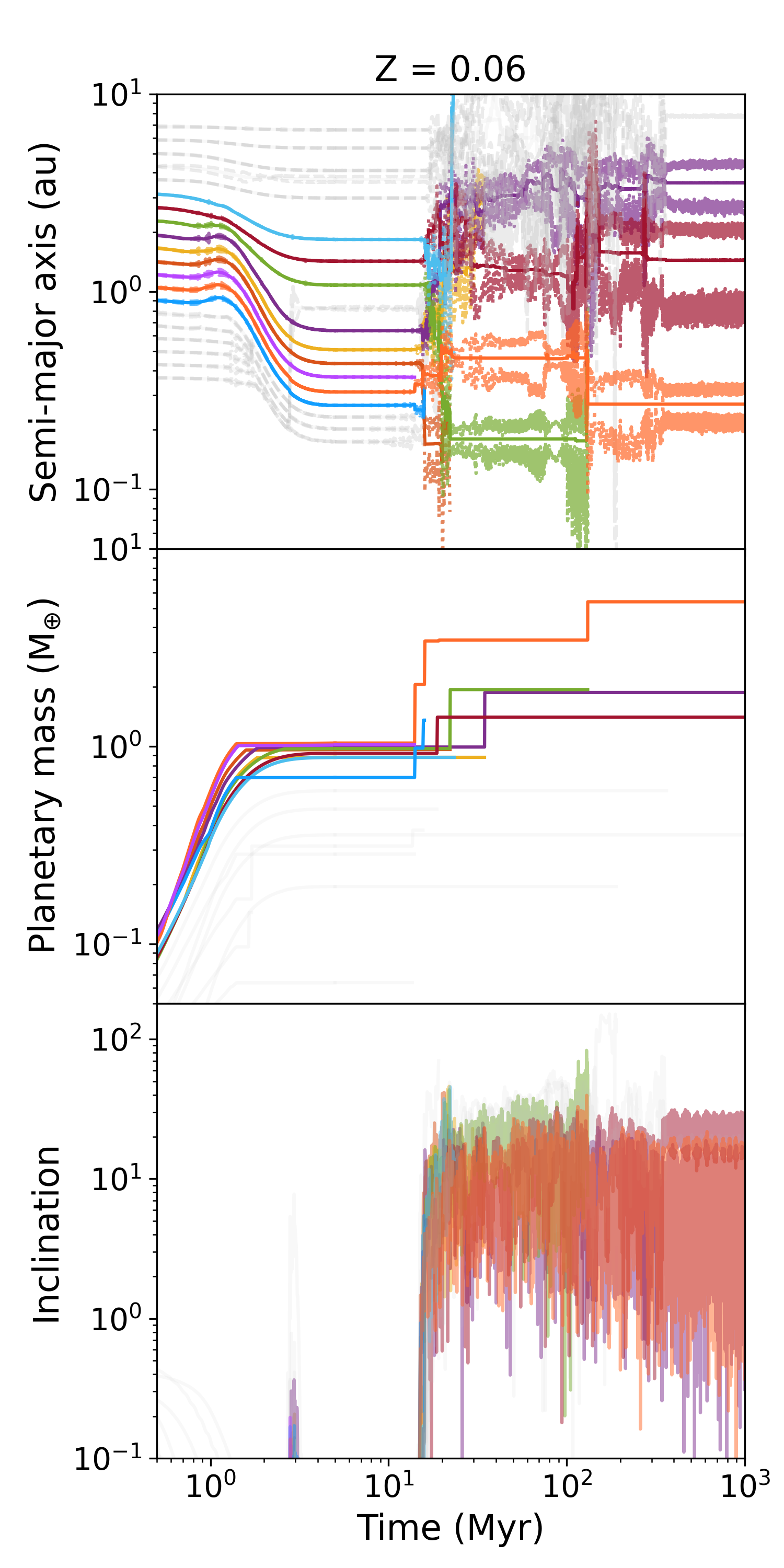}
    \caption{
    The growth and long-term evolution of a planetary system around a $0.3 \ M_{\odot}$ star. The initial disk accretion rate is $\dot{M}_{\rm g} = 6 \times 10^{-9} \ M_{\odot}/\rm{yr}^{-1}$, and the disk dissipates at approximately 5 Myr. The colorful curves represent planets with masses exceeding $0.7 \ M_{\oplus}$. The dotted lines in the upper panel indicate the perihelion and aphelion. Only one high-eccentricity, high-inclination, transiting single super-Earth remains, with an orbital period of less than 100 days.
     }
    \label{fig:longterm}
\end{figure}

We explain the pattern of eccentricity/inclination and stellar mass as follows. Multi-planet systems are natural outcomes of planet formation. During the initial gas disk phase, gas damping tends to reduce the relative motion of growing planets, resulting in near-circular and coplanar orbits. After gas disk dispersal, their eccentricities and mutual inclinations may be excited through long-term gravitational interactions. Once dynamical instabilities are triggered, collisions and ejections can further reduce the number of planets \citep{Izidoro2017}.   One example of the long-term evolution of a planetary system in stars of $Z_{\star}{=}0.06$ is shown in Fig.~\ref{fig:longterm}. Even if multiple planets ultimately survive in the system, their orbits often become highly misaligned, causing them to be observed as single planets.  Therefore, the single-transiting planets and their dynamical states ($e$ and $i$) are correlated. In metal-rich disks, the abundance of solids promotes the growth of more massive planets, which leads to more violent dynamical activities. As such, single-transit planets orbiting metal-rich stars tend to have higher eccentricities and inclinations compared to those around metal-poor stars. On the other hand, systems that have not experienced strong planet-planet scatterings remain stable, retaining relatively low eccentricities and inclinations.
 
\section{Conclusion} \label{sec:sum}

In this study, we have carried out the pebble-driven planet formation population synthetic model to investigate the occurrence rates and architecture of planet systems around stars from late M dwarf to solar-mass stars, with varying metallicities.  In addition to the fundamental processes included in our previous work \citep{Liu2019,Pan2024}, the updated model incorporates new features, such as stellar luminosity evolution, realistic pebble size due to drift and fragmentation, disk solid chemical compositions, and the sublimation of pebbles as they cross different ice lines.  Over $1,000$ N-body simulations have been conducted where their initial conditions are sampled based on a Monte Carlo approach.  


We summarize our key results as follows:
\begin{itemize}
    \item The occurrence rate of warm super-Earths increases with stellar mass from late to early M dwarfs, peaks at stars of $M_{\star}{\simeq} 0.5 M_{\odot}$ and then decreases as mass further increases towards the solar-type (Figure \ref{fig:Occ_vs_star}a). The first increase is because of the high solid mass around early M dwarfs compared to late M dwarfs. The decline afterward is due to the combined effects of larger disk size, slower planet migration, and more predominant influence of giant planets around GK stars compared to lower-mass dwarfs. In contrast, the occurrence rate of warm sub-giants, warm giants, and cold giants all exhibit a monotonic increase with stellar mass (Figures \ref{fig:Occ_vs_star}b,c and  \ref{fig:Occ_CJ})  
    
    \item The occurrence rate of gas giants strongly correlates with stellar metallicity. Nevertheless, the occurrence rate of small planets with $M_{\rm p}{=}1{-}20 \ M_{\oplus}$ exhibit a relatively weak $Z_{\star}$ dependency. In particular, $\eta_{\rm SE}$ is largely independent of   stellar metallicity for $Z_{\star}{>}-0.2$ (Figure \ref{fig:Occ_vs_Z}).

    \item We additionally simulate dozens of systems over $1$ Gyr and classify the resulting systems into ``observed" singles and multiples. In the ``observed" single-planet systems, the mean eccentricity and inclination increase significantly with stellar metallicity. This trend arises because massive planets are more likely to form in disks around metal-rich stars, potentially leading to frequent planet-planet scatterings. As a result, such systems tend to be dynamically hot. On the other hand, some multiple-planet systems can avoid dynamical instabilities, allowing the planets to maintain nearly circular and coplanar orbits 
      (Figure \ref{fig:ecc_inc}).
\end{itemize}

Overall, the results of our planet population synthesis are consistent with observational statistics, demonstrating that planet occurrence rates are significantly influenced by both stellar mass and metallicity. Furthermore, the analysis of long-term evolution reveals the correlations between the architecture of the system and stellar metallicity.  These findings advance our understanding of how stellar environment shapes the properties of planetary systems.

\section{Acknowledgments}
This work is supported by the National Key R\&D Program of China (2024YFA1611803). BL and MP are supported by the National Natural Science Foundation of China (Nos. 12222303,  12173035 and 12111530175), the start-up grant of the Bairen program from  Zhejiang University and the Fundamental Research Funds for the Central Universities (2022-KYY-506107- 0001,226-2022-00216).
J.-W. X. also acknowledges the support from the National Youth Talent Support Program
WZ acknowledges the National Natural Science Foundation of China (grant Nos. 12173021 and 12133005).
IR acknowledges further financial support from the European Research Council
(ERC) under the European Union’s Horizon Europe programme (ERC Advanced
Grant SPOTLESS; no. 101140786).
The simulations and analysis presented in this article were carried out on both and SilkRiver Supercomputer of Zhejiang University.


\appendix
\counterwithin{figure}{section}
\counterwithin{table}{section}

\section{Stellar luminosity evolution} \label{app:luminosity}

Planet formation occurs during the early stages of stellar evolution, when the evolving stellar luminosity significantly influences the thermal structure of protoplanetary disks. As the locations of the ice lines change over time, they affect the availability and distribution of condensed materials within the disk, thereby influencing the subsequent planet growth \citet{Panic&Min2017, Stammler2017, Miley2021}. Therefore, rather than using a fixed luminosity value, we incorporate the stellar luminosity evolution derived from \citet{Baraffe2015}. 

\begin{figure}
    \centering
    \includegraphics[width=\columnwidth]{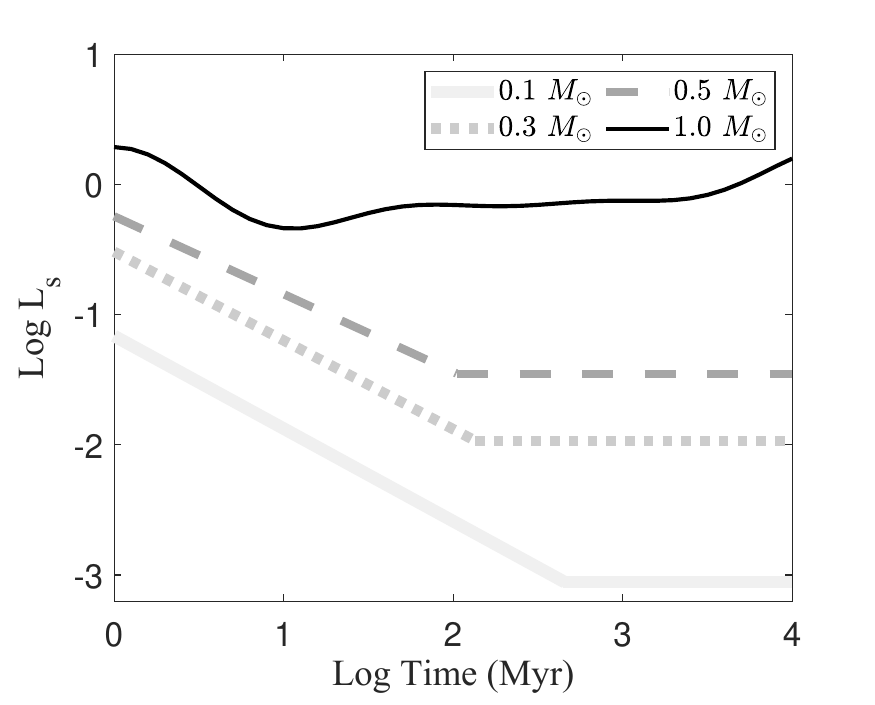}
    \caption{Time evolution of stellar luminosity based on \citet{Baraffe2015}'evolutionary model. The grey solid, dotted, dashed, and black solid lines correspond to stellar mass of $0.1$, $0.3$, $0.5$, and $1 \ M_{\odot}$, respectively.}
    \label{fig:Luminosity}
\end{figure}

The luminosity evolution for stars of different masses is illustrated in Fig.~\ref{fig:Luminosity}. For M dwarfs, the luminosity decreases monotonically as the star contracts. After $10^2$ to $10^3$ Myr, the stars enter the main sequence phase and their luminosities level off.  In contrast, solar-type stars undergo an initial phase of decreasing luminosity, followed by a brief increase before they enter the main-sequence phase. 

\section{Chemical composition} \label{app:chemical}

When pebbles drift across the ice line, the corresponding components sublimate. This process can significantly influence pebble accretion efficiency and the chemical composition of the accreting planets.

Following the conventional framework, we assume that the chemical abundances in protoplanetary disks match those of their host stars. We adopt the solar composition from \cite{Asplund2021} and focus on the partitioning of major elements.

Following the method of \citet{Schneider&Bitsch2021}, we simplify the chemical model by classifying all minerals as refractories. Given that  high abundance ($\sim 60\%$) of organic materials (including refractory organics and volatile organics) observed in comets \citep{Altwegg2019}, we include both organics ($\rm C_n H_x$) and carbon grain (C) in our model, assuming an equal proportion of each. 
The remaining 40\% of C is in the form of CO and $\rm CO_2$, with their relative abundances derived from \citet{Oberg2011}. Since $\rm C H_4$ and $\rm H_2 S$ account for no more than 1\% of the solid mass fraction,  their contribution is omitted. 

Our model includes a total of eight rocky or icy species: refractory materials, iron sulfide ($\rm FeS$), carbon grains ($\rm C$), water ($\rm H_2 O$), organics ($\rm C_n H_x$), carbon dioxide ($\rm CO_2$), carbon monoxide ($\rm CO$), and nitrogen ($\rm N_2$). The condensation temperatures of each species are taken from \citep{Lodders2003, Martin2014}. The volume mixing ratio of each species $v_{\rm Y}$ is depicted as a function of the relative abundance of element X to hydrogen H, and their corresponding mass fractions are listed in (Table~\ref{tab:species})

As pebbles cross ice lines, we assume that the corresponding volatile compounds fully sublimate with the Stokes number of the pebbles remains unchanged.


\begin{table*}[]
    \centering
    \caption{Condensation temperatures, volume mixing ratios, mass fractions, and cumulative mass fraction of the chemical species.}
    \begin{tabular}{ccccc}
        \hline
        \hline
        Species (Y) & $T_{cond}$ (K) & $v_{\rm Y}$ & $f_{mass}$ (\%) & $Cf_{mass}$\\
        \hline
        $\rm N_2$ & 20 & $0.45 \times \rm N/H^a$ & 8.0 & 1.000\\
        $\rm CO$ & 20 & $0.3 \times \rm C/H^{b, c}$ & 15.0 & 0.920\\
        $\rm CO_2$ & 70 & $0.1 \times \rm C/H^{b, c}$ & 7.5 & 0.770\\
        $\rm C_n H_x$ & 100 & $0.3 \times \rm C/H^a $ & 6.0 & 0.695\\
        $\rm H_2 O$ & 150 & $\rm O/H- 0.5 \times C/H-1.5 \times (0.8 \times Fe/H - S/H) - Mg/H - 0.2 \times Fe/H$ & \\
         & & $\rm - 1.5 \times Al/H - Ca/H - 0.5 \times Na/H - 2 \times Si/H$ & 22.0 & 0.635\\
        $\rm C$ & 631 & $0.3 \times \rm C/H^a$ & 6.0 & 0.415\\
        $\rm FeS$ & 704 & $\rm S/H^d$ & 6.5 & 0.355\\
        Refractories & 1500 & $\rm Mg/H + 0.6 \times Fe/H + 0.5 \times Al/H + Ca/H + 0.5 \times Na/H + Si/H + 0.5 \times S/H$ & 29.0 & 0.290\\
        \hline
    \end{tabular}
    \label{tab:species}
    \tablecomments{The condensation temperatures of different species are from \citep{Lodders2003, Martin2014}. Volume mixing ratios are based on the work of (a) \citet{Turrini2023}, (b) \citet{Pontoppidan2006}, (c) \citet{Oberg2011}, (d) \citet{Kama2019}. The mass fractions of each species are calculated according to the solar elemental abundances \citep{Asplund2021}. }
\end{table*}

\section{Planet mass-distance distribution} \label{app:M_a_diagram}

\begin{figure*} \label{fig:M_a}
    \centering
    \includegraphics[width=\textwidth]{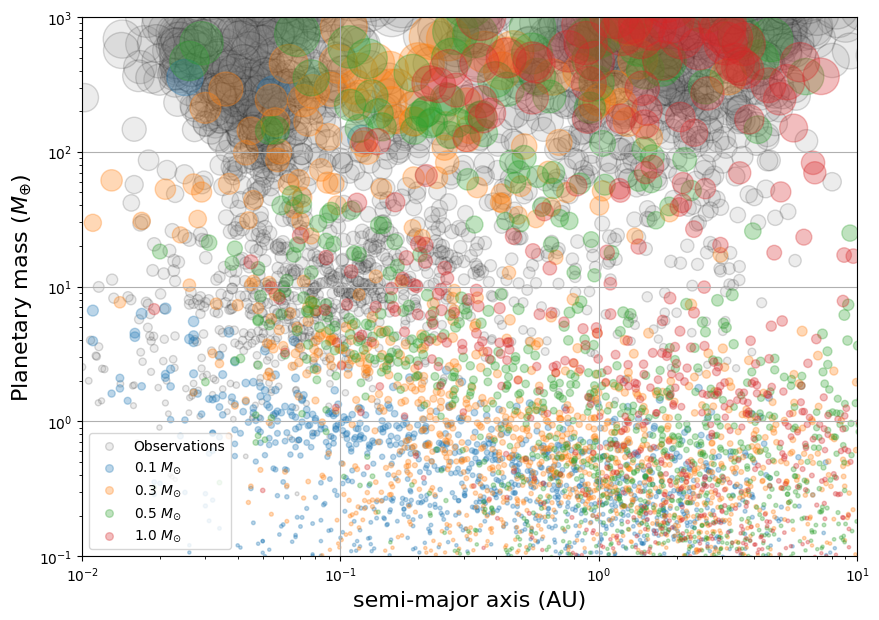}
    \caption{The mass-distance diagram of synthetic planets. The blue, orange, green, and red dots present planetary systems formed around $0.1$, $0.3$, $0.5$, and $1.0 \ M_{\odot}$ stars, respectively. Observations are indicated in grey. The size is scaled with planetary mass. 
    }
\end{figure*}

We present a comparison of simulated and observed planets in their mass and semi-major axis  diagram  in Fig. \ref{fig:M_a}. The observed and simulated populations are represented by gray and colored dots, and from red to blue refers to low to high stellar mass.

Due to observational biases, the exoplanet detection is sensitive to short-period and massive planets. Therefore, observations show a paucity of low-mass planets at wide orbits.

Our synthetic population reproduces several observational features, such as the prevalence of close-in small planets and the pile-up of cold Jupiters at $\sim 2{-3}$ au. Interestingly, the discrepancies still remain. For instance, our simulations produce a substantial number of long-period super-Earths and sub-Earths.  Due to the incompleteness of current surveys,  such populations is difficult to probe. We anticipate that future planet detection missions are needed to fill this gap and provide deep insights into the demographics of planetary systems \citep{Ge2022, Ji2022, Ji2024}.

\end{document}